\documentclass{elsart}
\usepackage{amsmath,amssymb,epsfig,subfigure}
\begin{document}
\runauthor{G. Ambika\thanks{ambika@iucaa.ernet.in} and K. Ambika}
\begin{frontmatter}
\title{Intermittency and Synchronisation in Gumowski-Mira Maps}
\author{G. Ambika}
\ead{ambika@iucaa.ernet.in}
\author{and K. Ambika}
\address{Department of Physics,
Maharaja's College, Cochin-682011, India.}

\begin{abstract}
The Gumowski-Mira map is a 2-dimensional recurrence relation that provide a
large variety of phase space plots resembling fractal patterns of nature. We
investigate the nature of the dynamical states that produce these patterns and 
find that they correspond to Type I intermittency near periodic cycles.
By coupling two GM maps, such patterns can be wiped out to give synchronised
periodic states of lower order. The efficiency of the coupling scheme is 
established by analysing the error function dynamics.
\end{abstract}
\end{frontmatter}

\section{Introduction}\label{Gu-sec1}
Non linear recurrence relations model many real world systems and help in
analysing their possible asymptotic behaviour as the parameters are 
varied~\cite{Gumo1}. Here we analyse one such recurrence map defined through
a 2-dimensional difference equation called Gumowski-Mira
transformation~\cite{Gumo2}. This system involves a non linear function $f(x)$
that is advanced in time by one iteration, in the $y$ equation. It has been
reported that computer simulations of this iterative scheme~\cite{Gumo3} give
rise to a variety of 2-dimensional images in the phase plane $(x,y)$ called
GM patterns that resemble natural objects like star fish, jelly fish, wings
of a butterfly, sections of fruits, flowers of varying shapes etc. Apart from 
a mere fascination in creating these patterns, one can implement a scheme like 
Iterative function scheme for producing 2-dimensional fractals or fractal
objects of nature using this map for  different values of the  parameters 
involved. Moreover, the final pattern is found to depend very sensitively on
the parameters, a feature which can be exploited in decision making algorithms
and control techniques in computing and communication.

One motivation in this work is to  analyse the underlying dynamics behind the 
formation of these patterns and their possible bifurcation sequences.
In this context, we would like to mention that the period doubling sequences and chaos doubling phenomenon
in this system has been reported earlier for a chosen set  of parameter 
values~\cite{Gumo4}. We are concentrating on another set 
that was used in~\cite{Gumo3} for generating GM patterns. Here the dynamics
involved is different in the sense that the 
intermittency route and odd period cycles predominate and the route to chaos
is via formation of quasiperiodic bands and band mergings. The detailed
evolution of the basin structure during the scenario until onset of chaos and 
final escape forms part of our work. The scaling behaviour near intermittency,
stability of the  lowest periodic cycles are also analysed. Moreover our work
provides evidence for periodic window and repeated substructures in the 
bifurcation scenario which show self similarity with respect to scaling in the
parameter. This can therefore account for the sensitive dependence of the 
asymptotic dynamical states or phase portraits on the relevant parameter of 
the system. 

It has been reported that~\cite{Gumo5} the lowest order periodic orbits of the
GM map are non parametric in nature and hence a straight forward application 
of OGY scheme to control chaos in the system is not desirable. Hence this is 
achieved through
a proxy system in~\cite{Gumo5}. We introduce a simple and straight forward
linear and mutual coupling scheme to achieve the same result. Moreover
by coupling two such  maps synchronised states with low periodic
cycles can be produced. Periodicity with synchronisity is found to be quite
robust in the system even against considerable perturbations.
A detailed analysis of the error function dynamics is carried out to estimate
the average synchronisation time $\langle\tau_1\rangle$ 
and the stabilisation time 
$\langle\tau_2\rangle$ after  a perturbation is applied. 

In this paper, Section 2 introduces the GM map and its possible stable 
dynamical states. Section 3 gives the results of a detailed numerical analysis
of the predominant bifurcation sequences in the map. The intermittency and 
other routes to chaos are given  in Section 4. In Section 5, the control of
chaos and synchronisation of the coupled maps is discussed. The discussion of 
the main results are given in Section 6. 

\section{Dynamical States of GM map.}\label{Gu-sec2}
\begin{figure}[t]
\centering
\mbox{
\subfigure[]{\epsfig{file=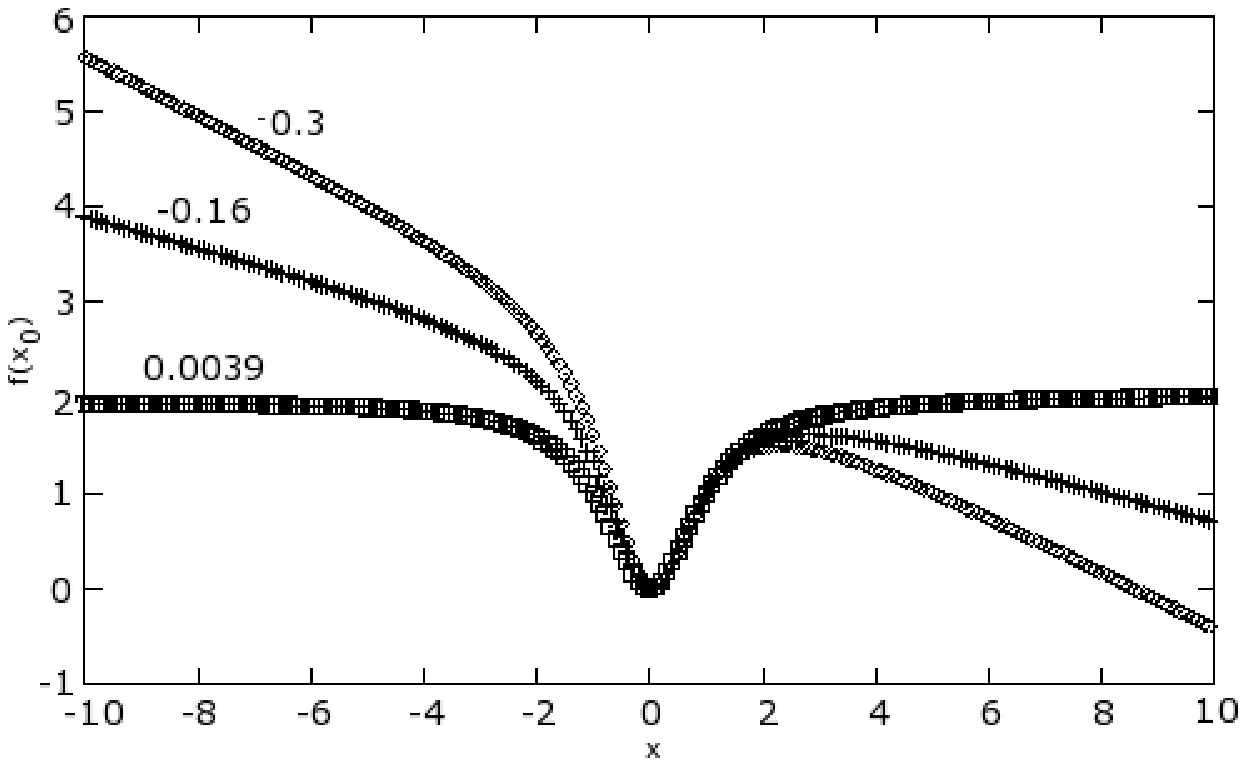,width=.5\linewidth}}
\subfigure[]{\epsfig{file=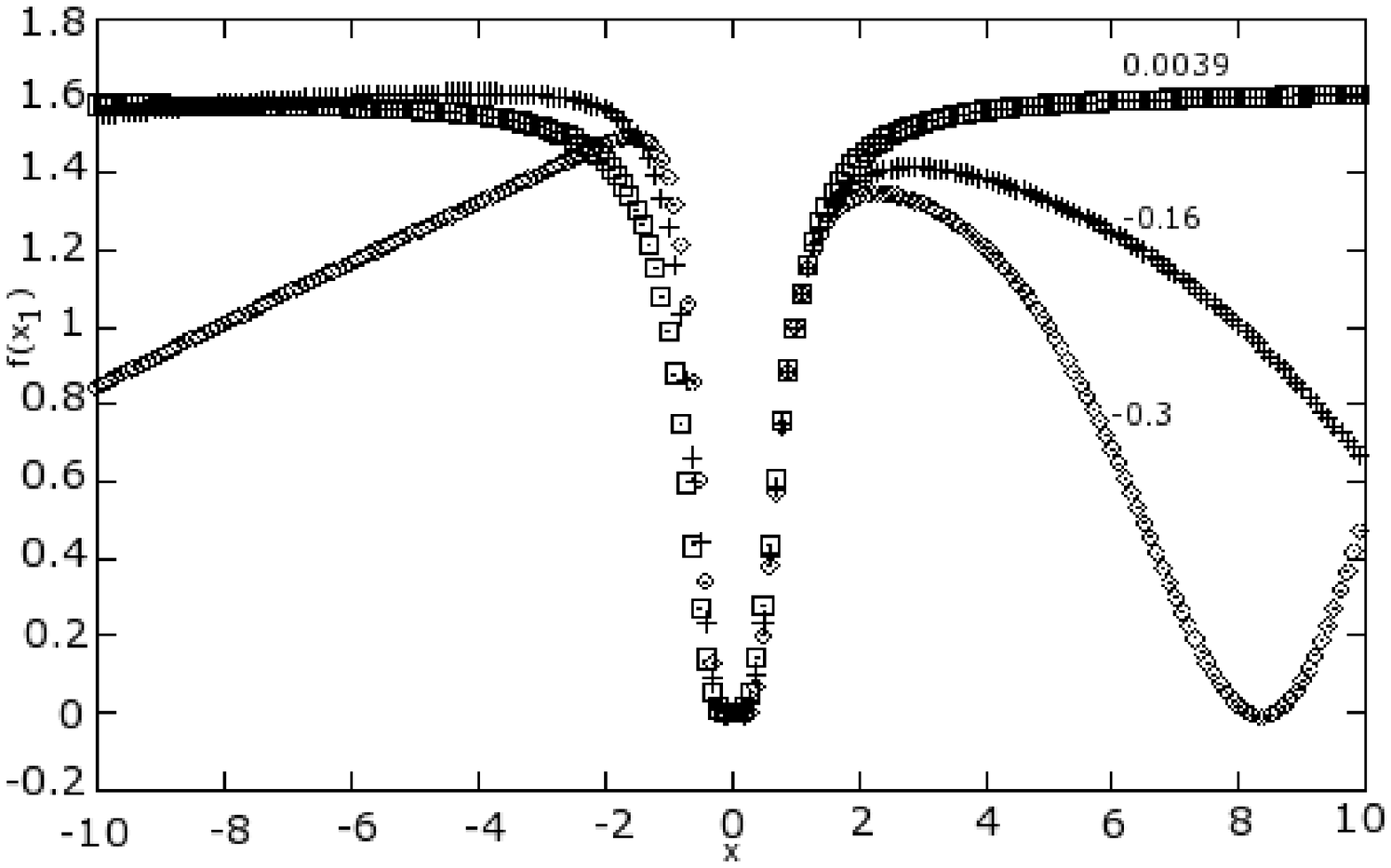,width=.5\linewidth}}}
\caption{Variation of a) the function $f(x)$ b) its first iterate for three
different values of $\mu$ $-0.3$, $-0.16$ and $0.0039$}\label{Gu-fig1}
\end{figure}

The transformation of Gumowski-Mira map that has been studied in computer 
simulations can be written as a 2-dimensional recurrence relation defined in
the $x-y$ plane as 

\begin{align}\label{Gu-eq2.1}
x_{n+1} &=y_n+a(1-by^2_n)y_n+f(x_n)\notag\\[5pt]
y_{n+1} &=-x_n+f(x_{n+1})
\end{align}
with $f(x_n)=\mu x_n+\frac{2(1-\mu)}{1+x^2_n}x^2_n$. It should be  noted that 
the function $f(x)$ in the $y$ equation is advanced in time by one iteration 
which provides richer and complex dynamical behaviour. The nature of variation
of the non linear function $f(x_0)$ as well as $f(x_1)$ for different $\mu$ 
values in the range $-0.3$ to $+0.0039$ with $a=0.008$ and $b=0.05$ is shown 
in figs~\ref{Gu-fig1}a and \ref{Gu-fig1}b. Here $x_0$ is the starting seed 
value and $x_1$ is the first iterate. These figures  
give an indication of how different the functions in the $x$ and $y$
equations are and how fast and 
differently they change as iterations proceed.
As any stable periodic cycle is reached asymptotically, these functions are
found to level off.

The one cycle fixed points of the system is determined by solving the equations
\begin{align}\label{Gu-eq2.2}
x&=y+a(1-by^2)y+\mu x+\frac{2(1-\mu)x^2}{1+x^2}\notag\\[5pt]
y&=(\mu-1)x+\frac{2(1-\mu)x^2}{1+x^2}
\end{align}
using MAXIMA and the only real and bounded solutions of interest are 
$(0,0)$ and $(1,0)$, 
which are unstable in the relevant parameter regions of our study. 
The 
most prominent elementary cycle here is a 4 cycle born by saddle node 
bifurcation. The stability
of this 4 cycle can be  established numerically  by calculating the eigen 
values of the Jacobian M for chosen values of $\mu$ and the corresponding 
elements of the 4 cycle. 
Thus if ($\delta x_n$, $\delta y_n$) are the small perturbations to the cycle
elements, we have
\begin{equation}\label{Gu-eq2.3}
\begin{pmatrix}\delta x_{n+1}\\\delta y_{n+1}\end{pmatrix}=M
\begin{pmatrix}\delta x_{n}\\\delta y_{n}\end{pmatrix}.
\end{equation}
where
\begin{align}\label{Gu-eq2.4}
M_{11} =& -\frac{4x^3(1-\mu)}{(1+x^2)^2}+\frac{4x(1-\mu)}{1+x^2}+\mu\notag\\
M_{12} =& 1-2aby^2+a(1-by^2)\notag\\
M_{21} =&-1+\mu\left(-\frac{4x^3(1-\mu)}{(1+x^2)^2}+\frac{4x(1-\mu)}
          {1+x^2}+\mu\right)-\notag\\
        &\frac{4(1-\mu)\left(-\frac{4x^3(1-\mu)}{(1+x^2)^2}+\frac{4x(1-\mu)}
          {1+x^2}+\mu\right)\left(y+ay(1-by^2)+\frac{2x^2(1-\mu)}{1+x^2}
          +x\mu\right)^3}{\left(1+(y+ay(1-by^2)+\frac{2x^2(1-\mu)}{1+x^2}
          +x\mu )^2\right)^2}+\notag\\
        & \frac{4(1-\mu)\left(-\frac{4x^3(1-\mu)}{(1+x^2)^2}+\frac{4x(1-\mu)}
          {1+x^2}+\mu\right)\left(y+ay(1-by^2)+\frac{2x^2(1-\mu)}{1+x^2}
          +x\mu\right)}{1+(y+ay(1-by^2)+\frac{2x^2(1-\mu)}{1+x^2}
          +x\mu )^2}
\end{align}
\begin{align*}
M_{22} =& (1-2aby^2+a(1-by^2))\mu-\notag\\
        & \frac{4(1-2aby^2+a(1-by^2))(1-\mu)(y+ay(1-by^2)+\frac{2x^2(1-\mu)}
          {1-x^2}+x\mu)^3}{\left(1+(y+ay(1-by^2)+\frac{2x^2(1-\mu)}{1-x^2}
          +x\mu )^3\right)^2}+\notag\\
        & \frac{4(1-2aby^2+a(1-by^2))(1-\mu)(y+ay(1-by^2)+\frac{2x^2(1-\mu)}
          {1+x^2}+x\mu)}{1+(y+ay(1-by^2)+\frac{2x^2(1-\mu)}{1+x^2}
          +x\mu )^2}\notag
\end{align*}
The eigen values of $M$ evaluated for a few typical $\mu$ values with 
$a=0.008$; $b=0.05$ in the 4 cycle window are given in~\ref{Gu-tab1} .

\begin{table}[h]
\centering
\caption{The Eigen values of the stability matrix for different $\mu$ values
in the 4 cycle window $|E|<1$ in all the above cases indicating stability of 
the concerned 4 cycle~\cite{Gumo6}}\label{Gu-tab1}
\begin{tabular}{|l|l|l|}
\hline
$\mu$      & $E_1$    & $E_2$\\\hline
$-0.0964$  & $0.6384$ & $-0.6346$\\\hline
$-0.095$   & $0.6451$ & $-0.6415$\\\hline
$-0.005$   & $0.9371$ & $-0.9371$\\\hline
$0.0002$   & $0.8169$ & $-0.8169$\\\hline
$0.0003$   & $0.7841$ & $-0.7841$\\\hline
\end{tabular}
\end{table}

\section{Bifurcation Sequence.}\label{Gu-sec3}
\begin{figure}[b]
\centering
\mbox{
\subfigure[]{\epsfig{file=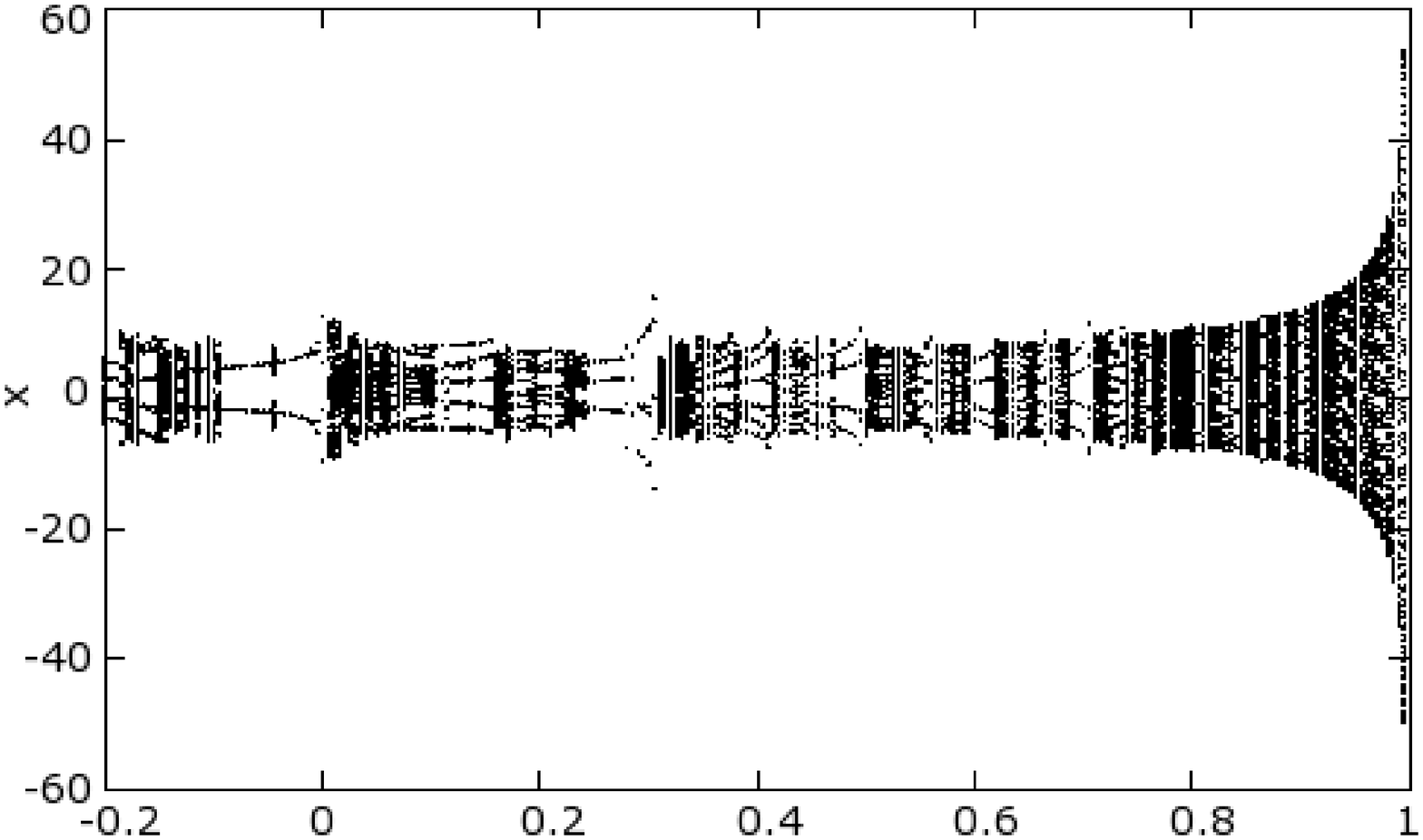,width=.5\columnwidth}}
\subfigure[$-0.3<\mu<-0.2005$]{\epsfig{file=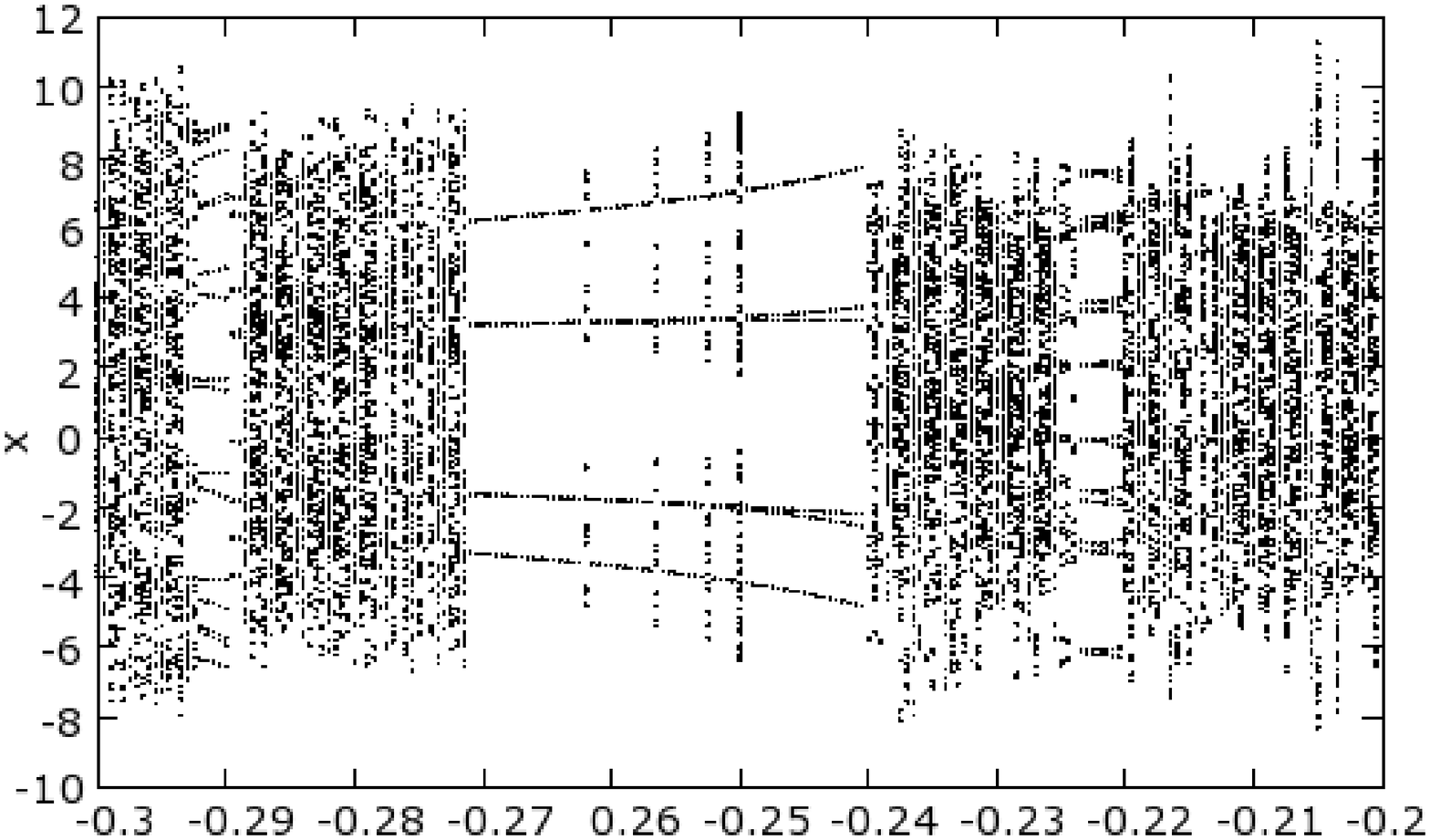,width=.5\columnwidth}}}\\
\mbox{
\subfigure[$-0.2<\mu<-0.0505$]{\epsfig{file=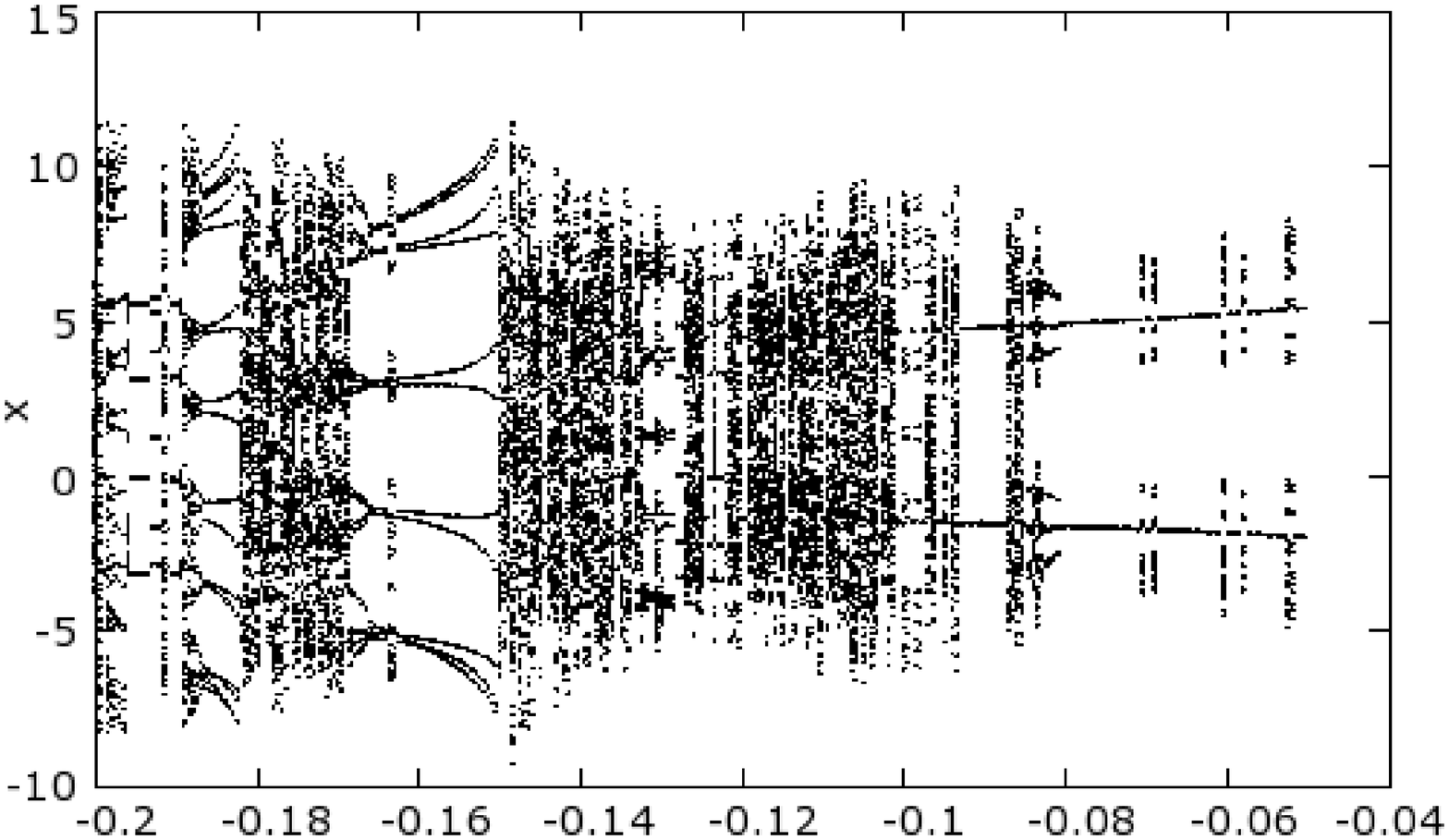,width=.5\columnwidth}}
\subfigure[$-0.01<\mu<-0.00095$]{\epsfig{file=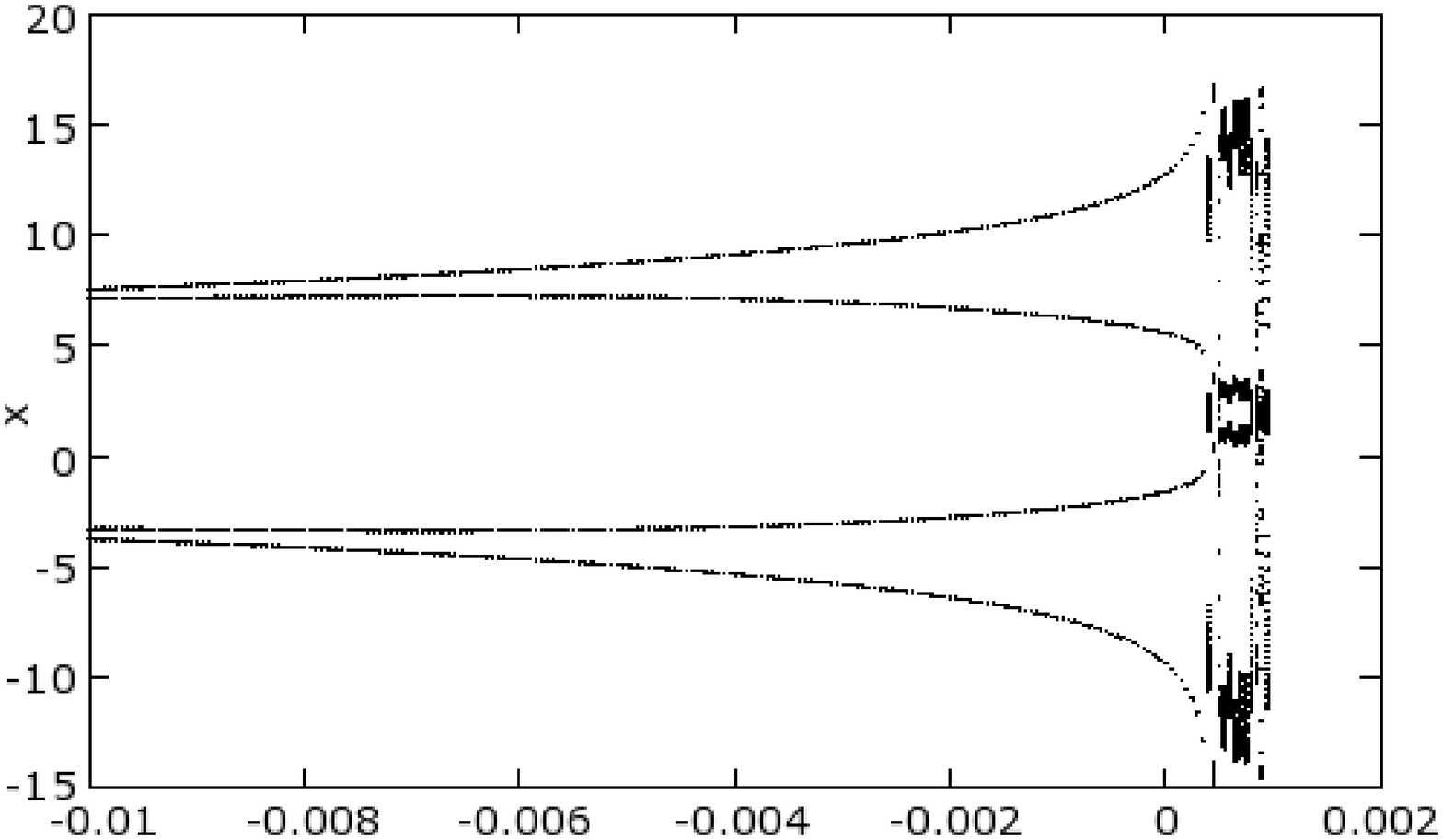,width=.5\columnwidth}}}
\caption{a) Bifurcation scenario in the range $\mu=-0.2$ to $+1$, for 
$a=0.008$, $b=0.05$. The windows of periodic cycles are zoomed for details in
b, c and d as indicated.}\label{Gu-fig2}
\end{figure}

We have mentioned that the asymptotic dynamical states of the GM map depends
sensitively on the parameter $\mu$. In this section, we numerically 
investigate in detail, the possible states and their bifurcation patterns as 
$\mu$ is varied. We find that the bounded interval for the map lies in the 
interval $[-1,1]$ for $a=0.008$ and $b=0.05$.
Moreover the most prominent elementary cycle of periodic behaviour is 4 which occurs
in many intermittent windows of $\mu$ in the bifurcation diagram. The full 
scenario is given in Fig~\ref{Gu-fig2}a which is mostly dominated by broad
windows of odd cycles like 7, 11 etc. Here out of 10,000 iterate 9000 are
discarded as initial transients and the next 1000 are plotted. 
The specific regions of 
the windows of such cycles are zoomed and reproduced in Fig~\ref{Gu-fig2}b,
\ref{Gu-fig2}c and~\ref{Gu-fig2}d.

\begin{figure}[t]
\centering
\mbox{
\subfigure[]{\epsfig{file=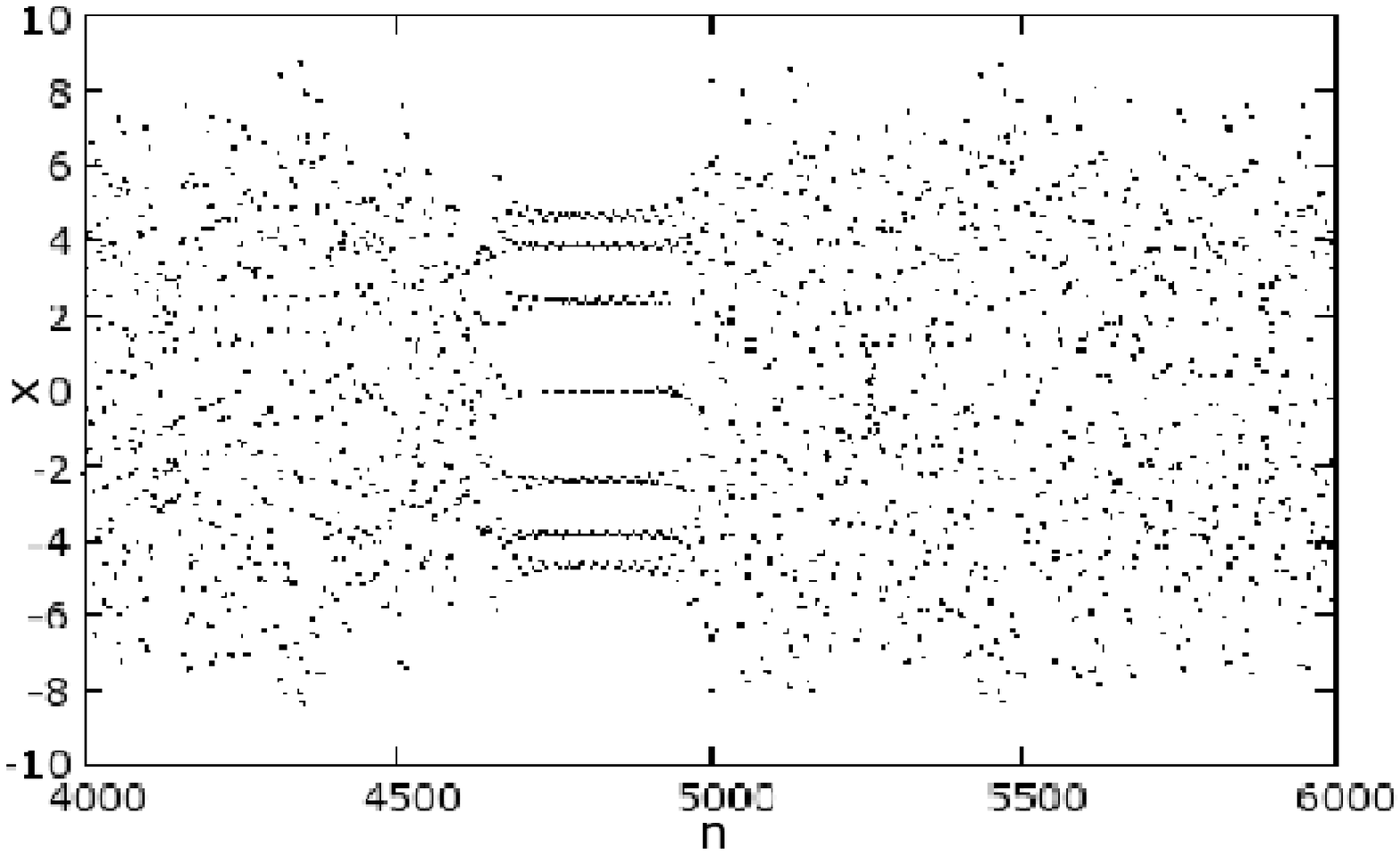,width=.5\columnwidth}}
\subfigure[]{\epsfig{file=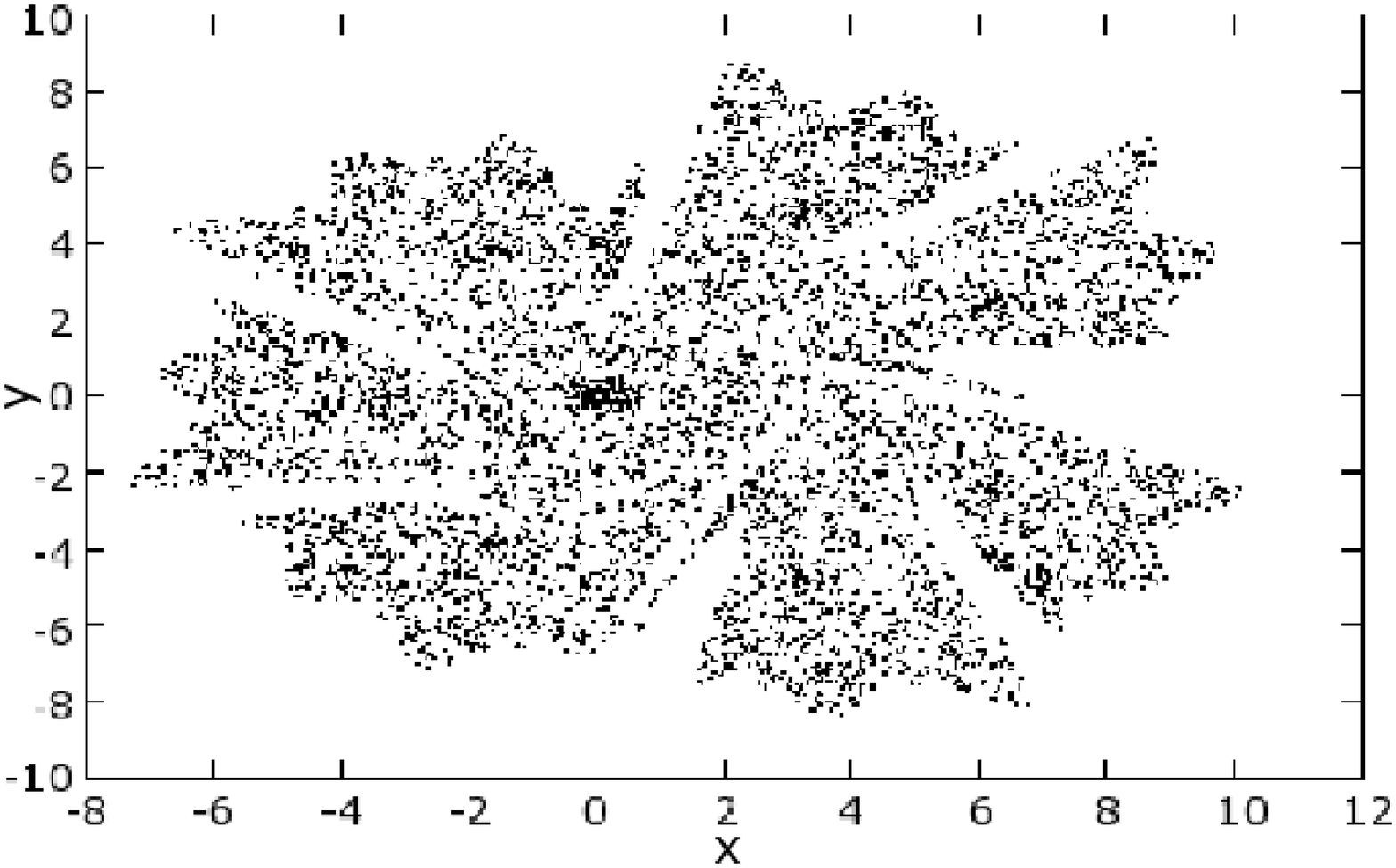,width=.5\columnwidth}}}\\
\mbox{
\subfigure[]{\epsfig{file=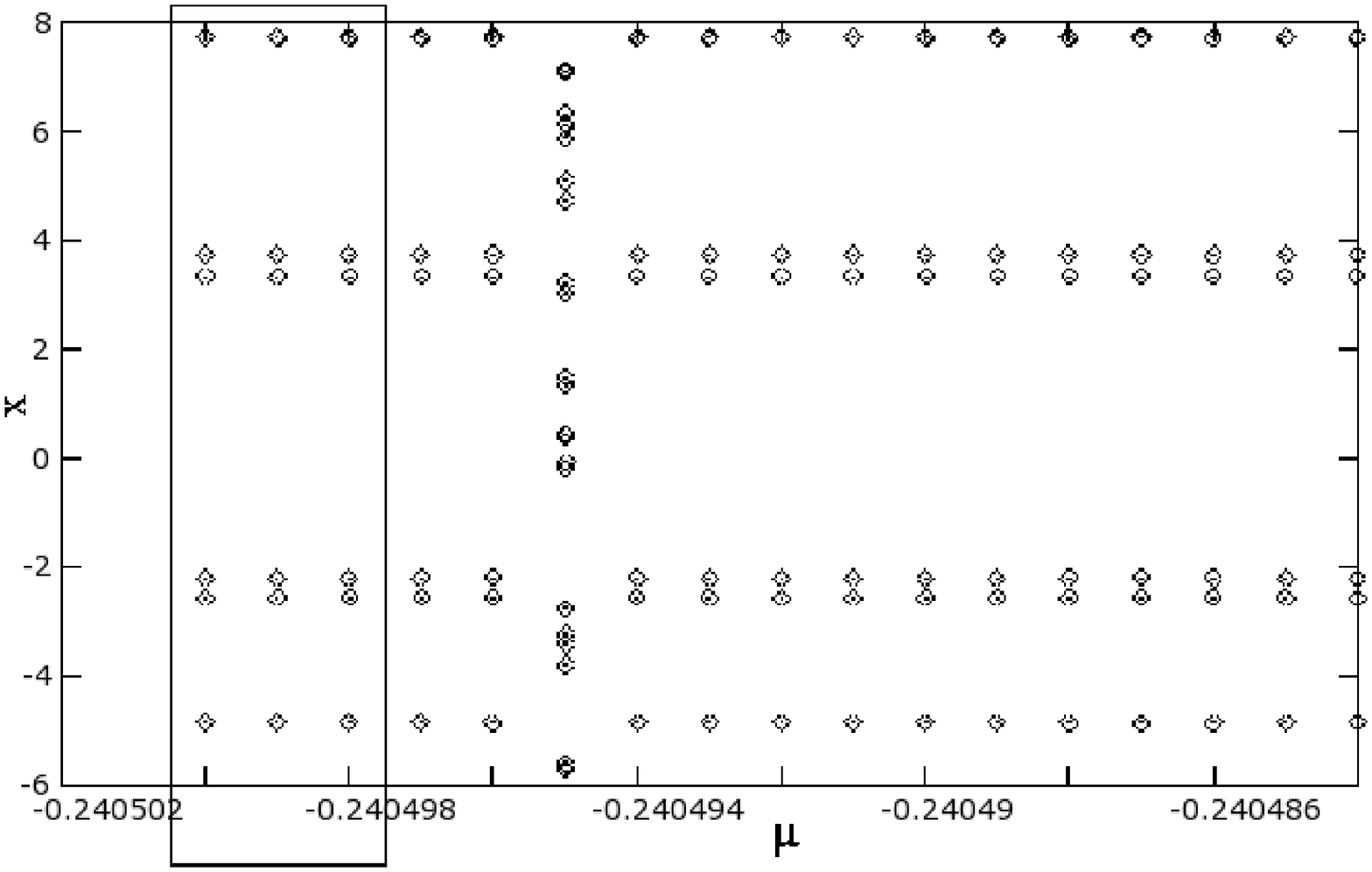,width=.5\columnwidth}}
\subfigure[]{\epsfig{file=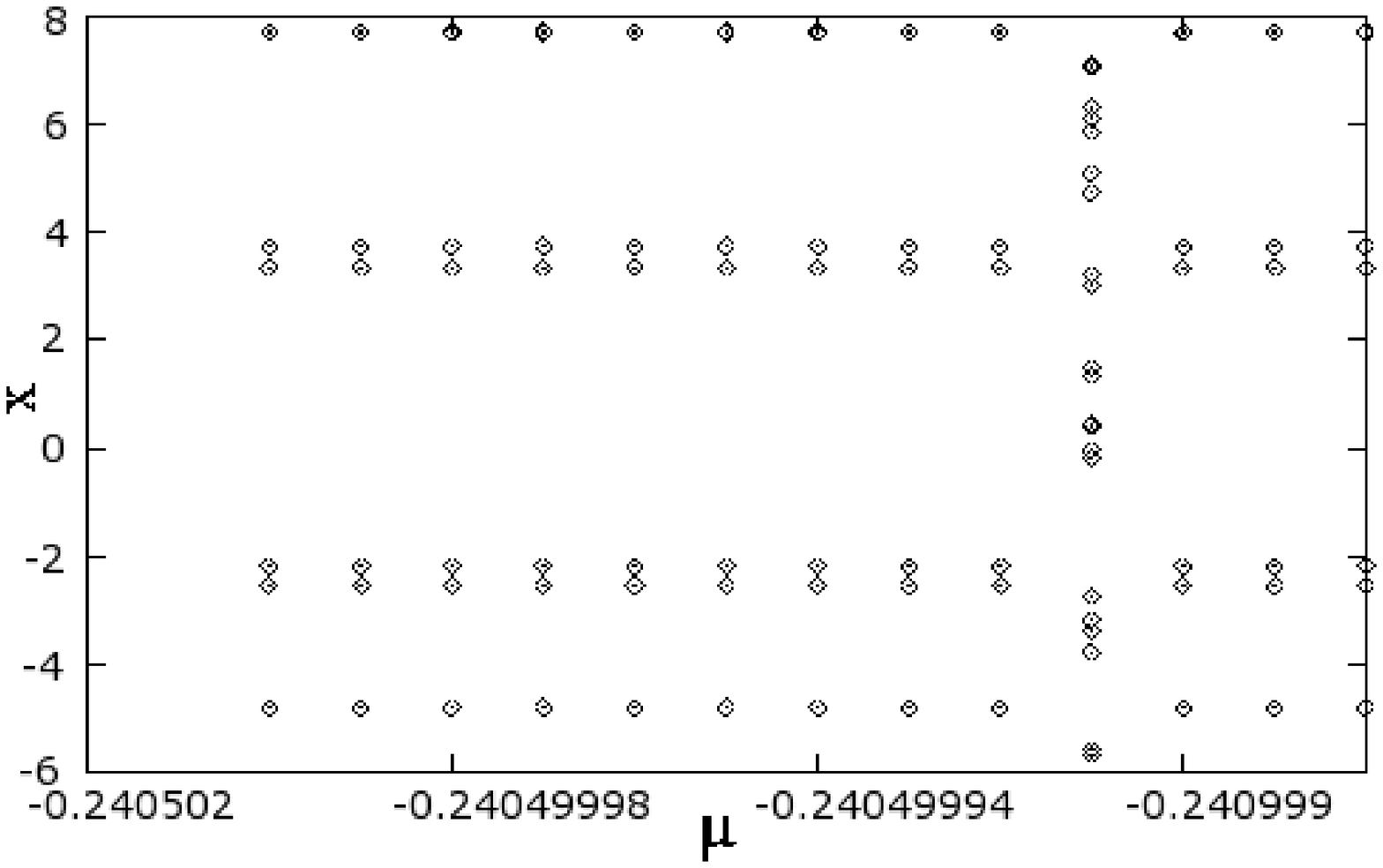,width=.5\columnwidth}}}
\caption{The intermittency behaviour before the stabilisation of a 7 cycle
is shown in the $x_n$-$n$ plot in (a) and corresponding phase portrait
in (b) for $\mu=-0.2734$. Note that this is the typical behaviour near the 
onset of each periodic window. The self similar and repeating substructures
inside this stability window is shown in Figs~\ref{Gu-fig3}c and 
\ref{Gu-fig3}d where  the 7 and 22 cycles are seen to recurr. (The
rectangle shown in \ref{Gu-fig3}c is zoomed in~\ref{Gu-fig3}d).}\label{Gu-fig3}
\end{figure}

In general these windows of periodic cycles 
born by tangent bifurcations or saddle node bifurcations at
their left ends exhibit intermittency behaviour in their iterates.
The transition to chaos takes place for small increase of values of $\mu$ when
the periodic cycle becomes unstable giving rise to quasiperiodic bands which
become chaotic and merge together. We illustrate this for the 7 cycle window.
Near the left end of a periodic 
7 cycle window Fig~\ref{Gu-fig3}a gives the 
$x_n-n$ plot with $\mu=-0.2734$ for 6000 iterations showing intermittently
laminar and chaotic behaviour.
Fig~\ref{Gu-fig3}b gives the corresponding $x-y$ plot.
It is interesting  to note that the periodic
window structure shows self similarity in their substructures with respect to 
scaling in $\mu$. This 
is evident from the plots in Fig~\ref{Gu-fig3}c and Fig~\ref{Gu-fig3}d where
a 7 cycle and a 22 cycle patterns repeat at two levels of zooming in the 
parameter range. This recurring  self similarity with the consequent 
intermittency in the beginning of each periodic cycle and quasi-periodicity at
its end, accounts for the variety and richness of the GM patterns in the 
$x-y$ plane with its highly sensitive dependence on $\mu$.

\section{Intermittency \& Transition to Chaos.}\label{Gu-sec4}
\begin{figure}[b]
\centering
\mbox{
\subfigure[]{\epsfig{file=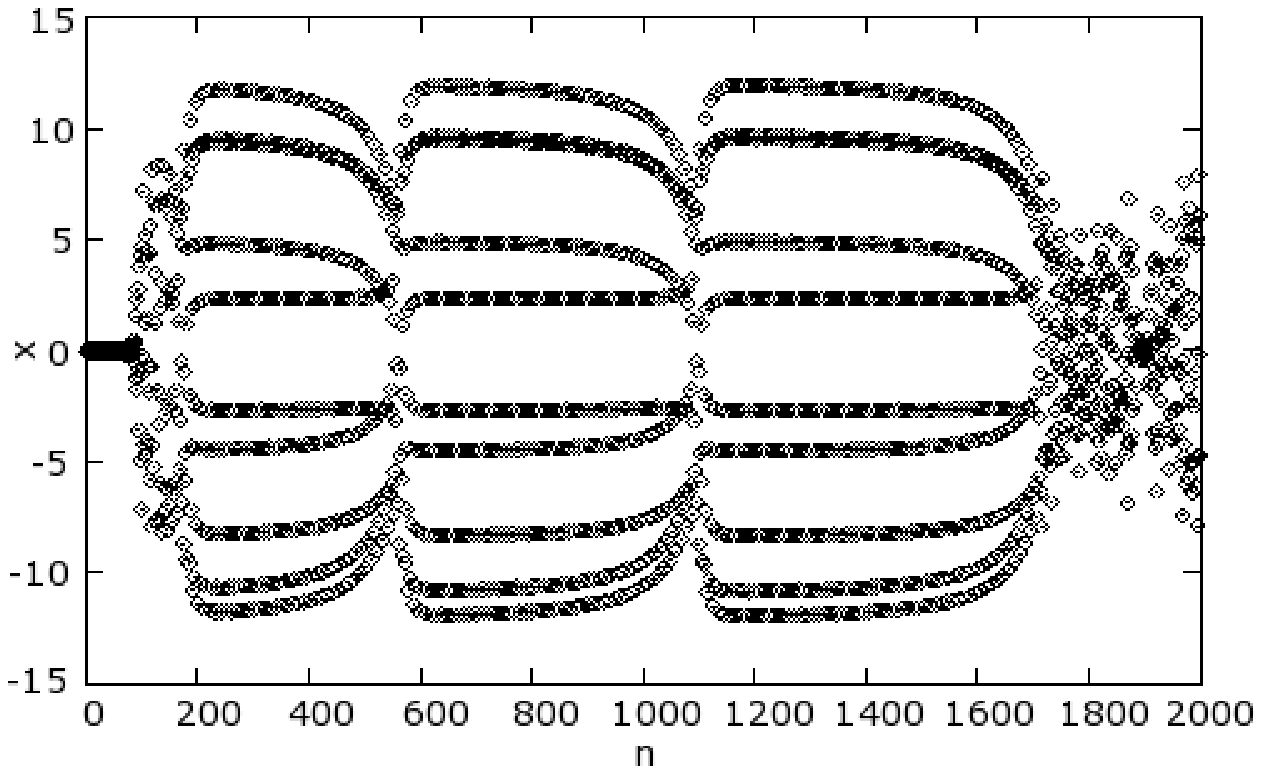,width=.5\columnwidth}}
\subfigure[]{\epsfig{file=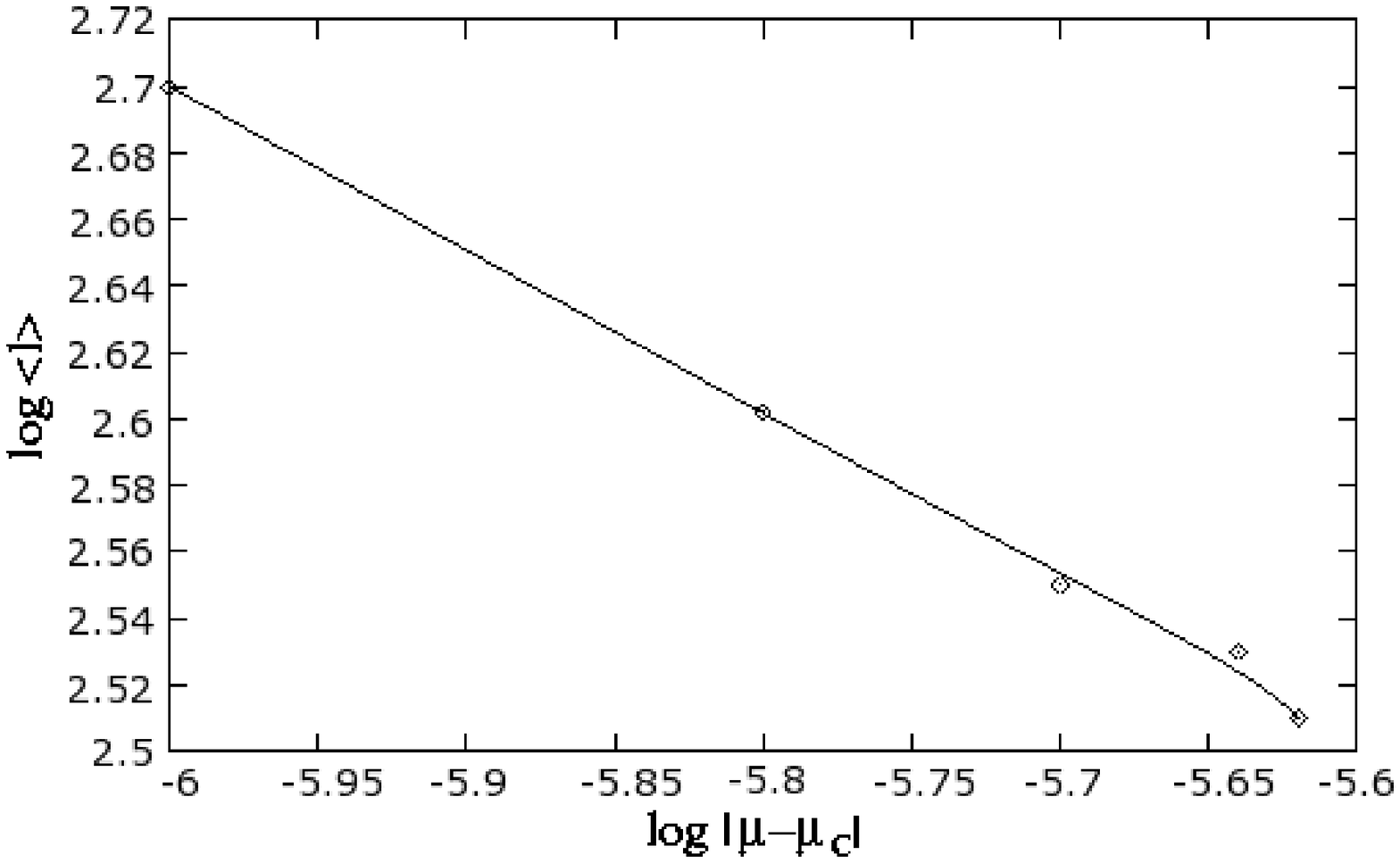,width=.5\columnwidth}}}
\caption{Type I intermittency near the onset of a 10 cycle a) the laminar 
and irregular behaviour for $\mu=-0.312501$ b) the scaling of average laminar 
region $\langle l \rangle$ as function of $|\mu-\mu_c|$
where $\mu_c=-0.312498$. The scaling index in this case is 
$-0.49$.}\label{Gu-fig4}
\end{figure}

The nature of intermittency before the birth of a periodic cycle is analysed 
by calculating the average life time of the laminar
region $\langle l\rangle$ in the $x_{n}-n$ plots. 
For one such region near $\mu=-0.312501$
before a ten cycle window the variation of $\langle l\rangle$ as a function of 
$|\mu-\mu_c|$ is studied (Fig~\ref{Gu-fig4}). 
$\mu_c$ is the critical value when it becomes chaotic. 
We can write $\langle l\rangle \sim |\mu-\mu_c|^\nu$. $\nu$ then
defines the scaling index that helps to identify the type of intermittency.
Here $\mu_c=-0.312498$ and $\nu=-0.49$ so that the intermittency is 
Type I~\cite{Gumo7} which is usually associated with saddle node bifurcation.

For further work we concentrate mainly on the stable window of the 4 cycle in
the range $-0.096<\mu<0.0003695$.

\begin{figure}[t]
\centering
\mbox{
\subfigure[]{\epsfig{file=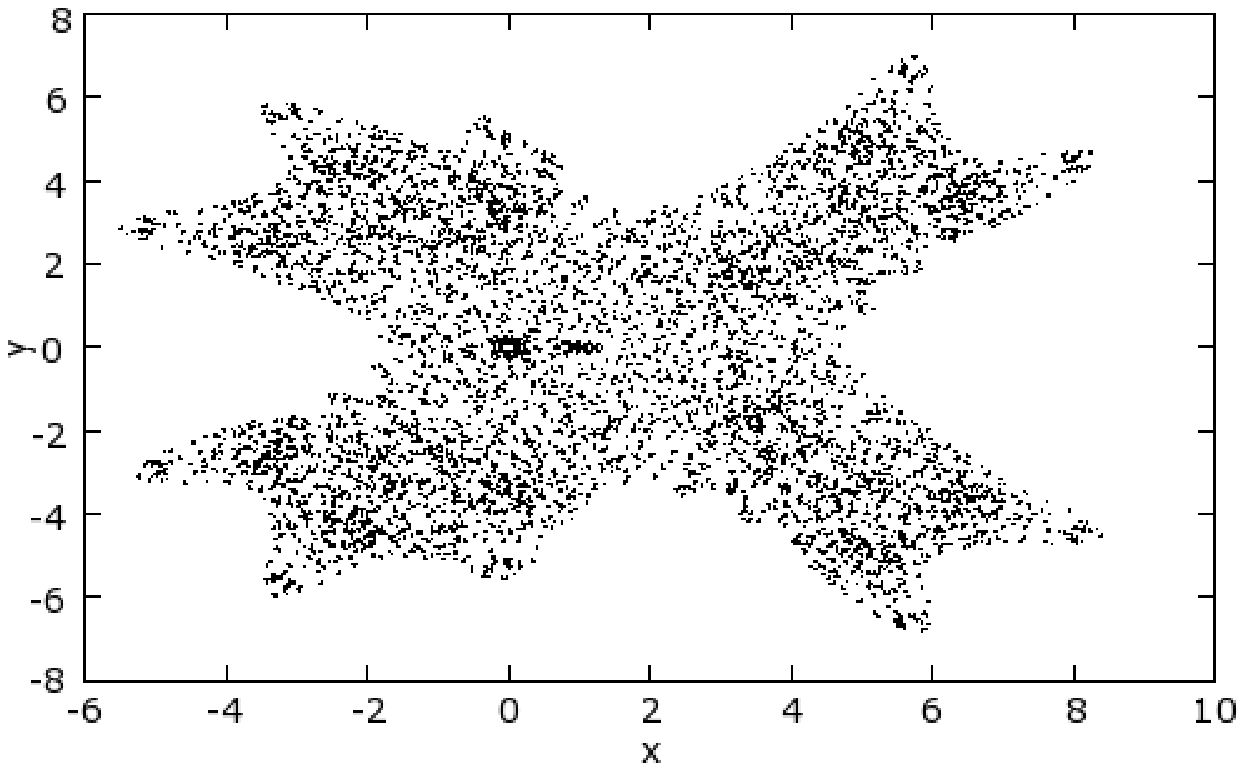,width=.5\columnwidth}}
\subfigure[]{\epsfig{file=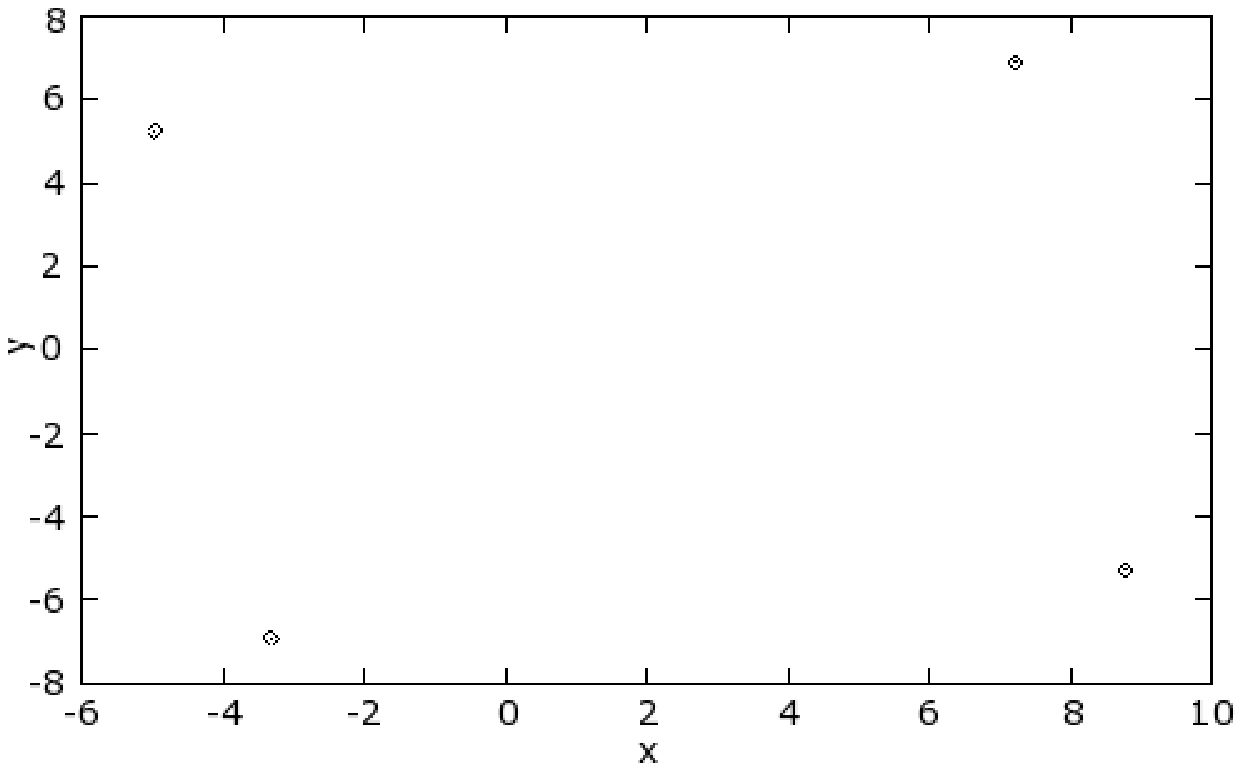,width=.5\columnwidth}}}\\
\mbox{
\subfigure[]{\epsfig{file=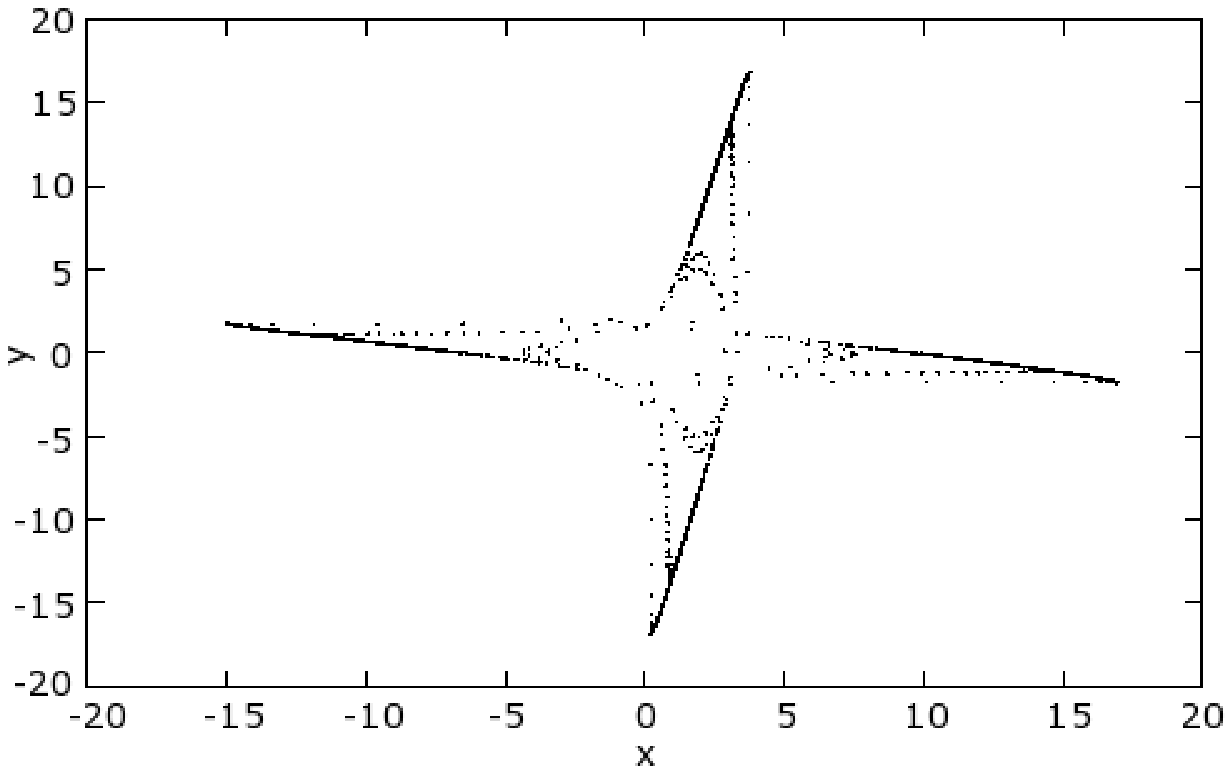,width=.5\columnwidth}}
\subfigure[]{\epsfig{file=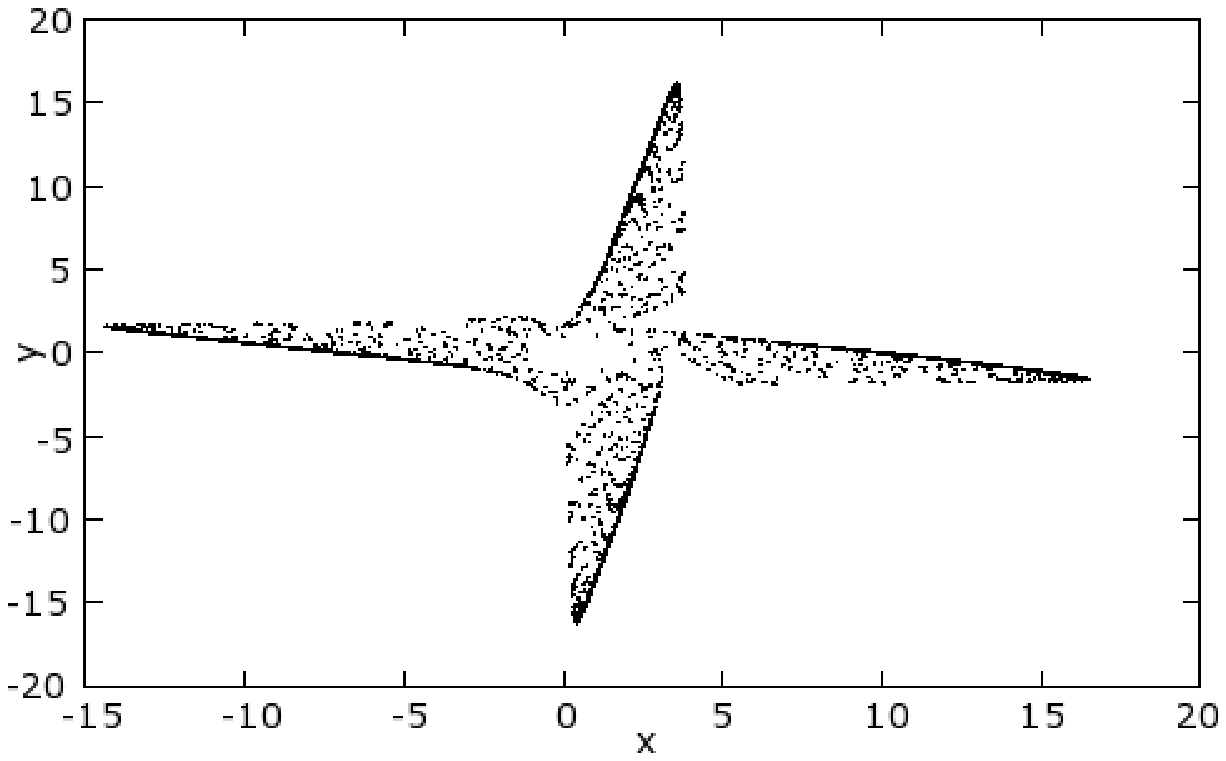,width=.5\columnwidth}
}}
\caption{The phase portraits for $\mu$ values in the region of the 4 cycle 
window
a) intermittency region ($\mu=-0.1199$)
b) stable 4 cycle ($\mu=-0.005$)
c) quasiperiodic 4 lands ($\mu=0.0003695$) and
d) chaotic attractor ($\mu=0.001$)}\label{Gu-fig5}
\end{figure}

The phase space during intermittency before the stabilisation of the 4 cycle,
the 4 cycle region, the quasi periodic band region with 4 bands and final
merging of  bands to chaos are shown in Fig~\ref{Gu-fig5}a--d. It is clear
that the transition to chaos is via quasi-periodicity to chaotic bands and 
band merging. The precise transition point is in this case $\mu=0.000369$.

\begin{figure}[t]
\centering
\mbox{\subfigure[]{\epsfig{file=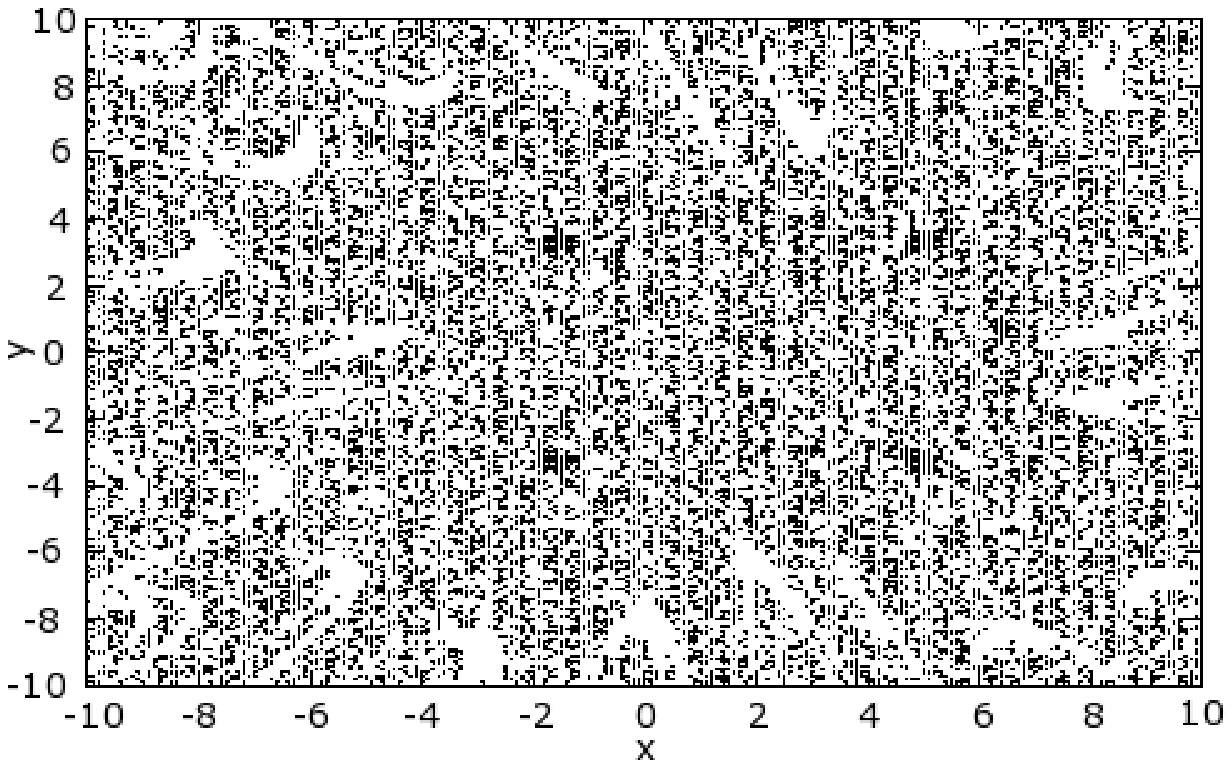,width=.5\columnwidth}}
\subfigure[]{\epsfig{file=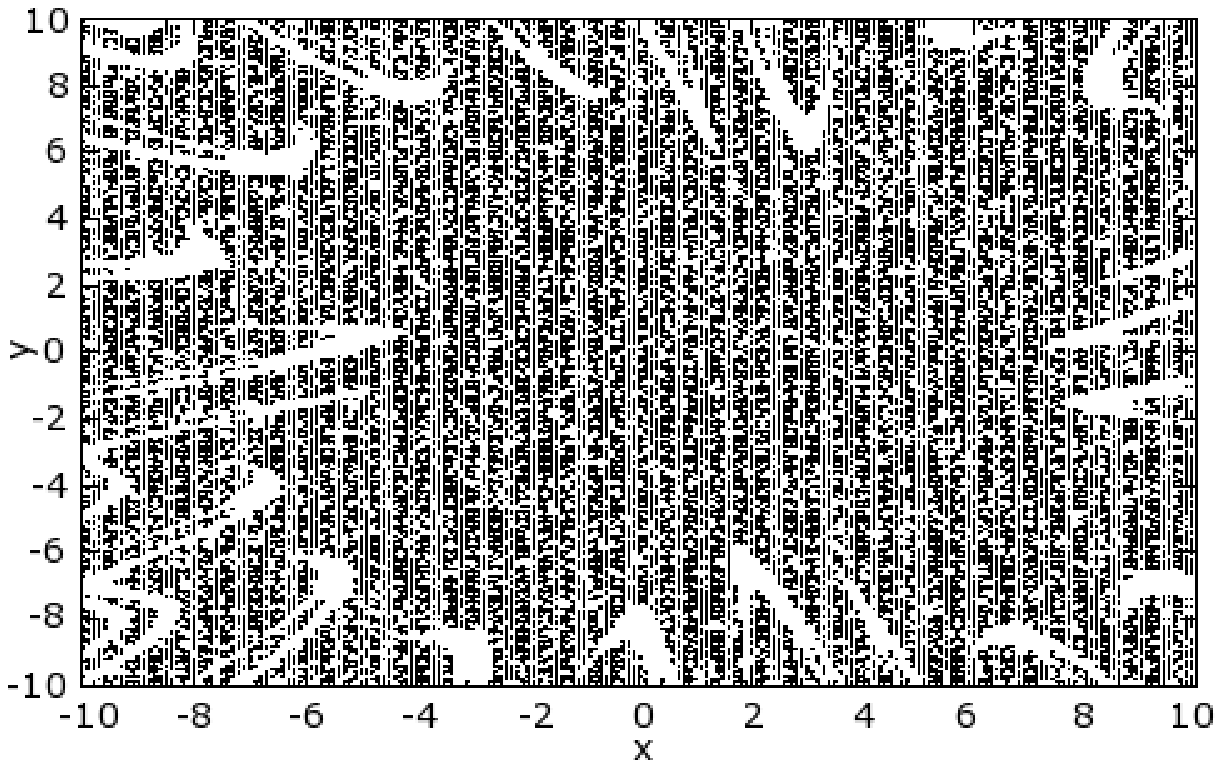,width=.5\columnwidth}}}\\
\mbox{\subfigure[]{\epsfig{file=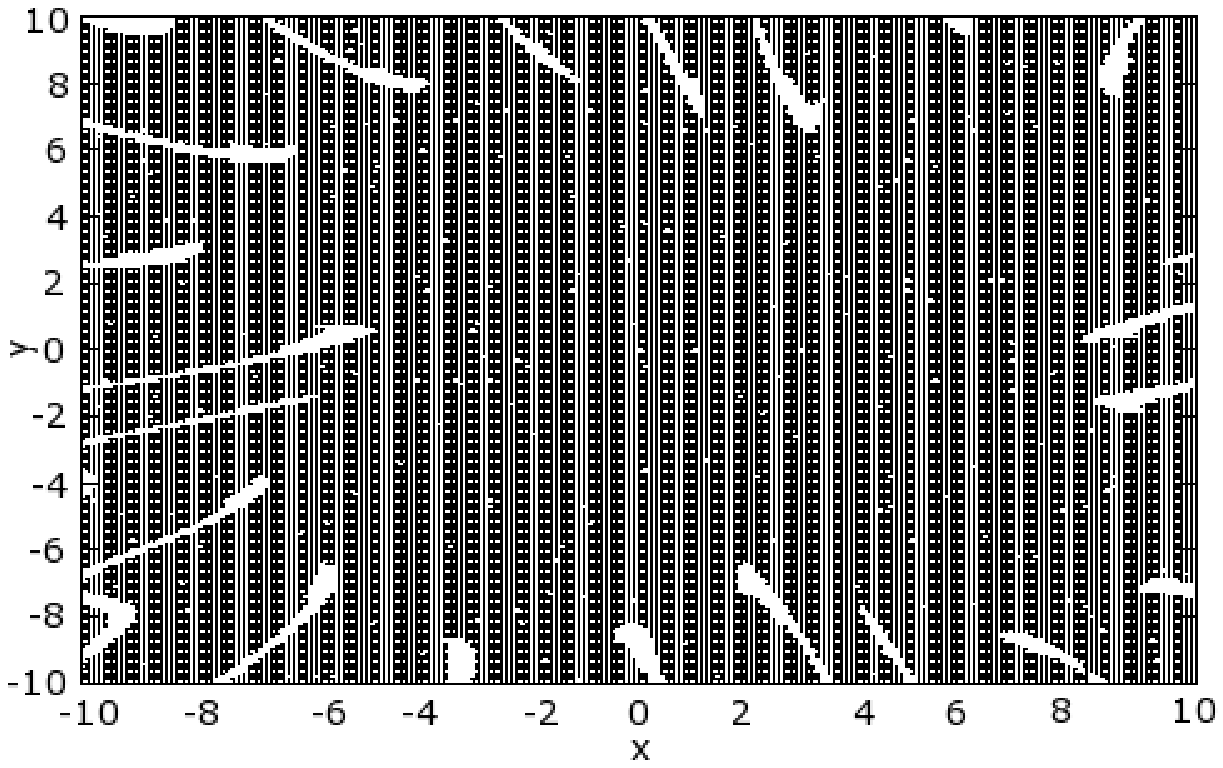,width=.5\columnwidth}}
\subfigure[]{\epsfig{file=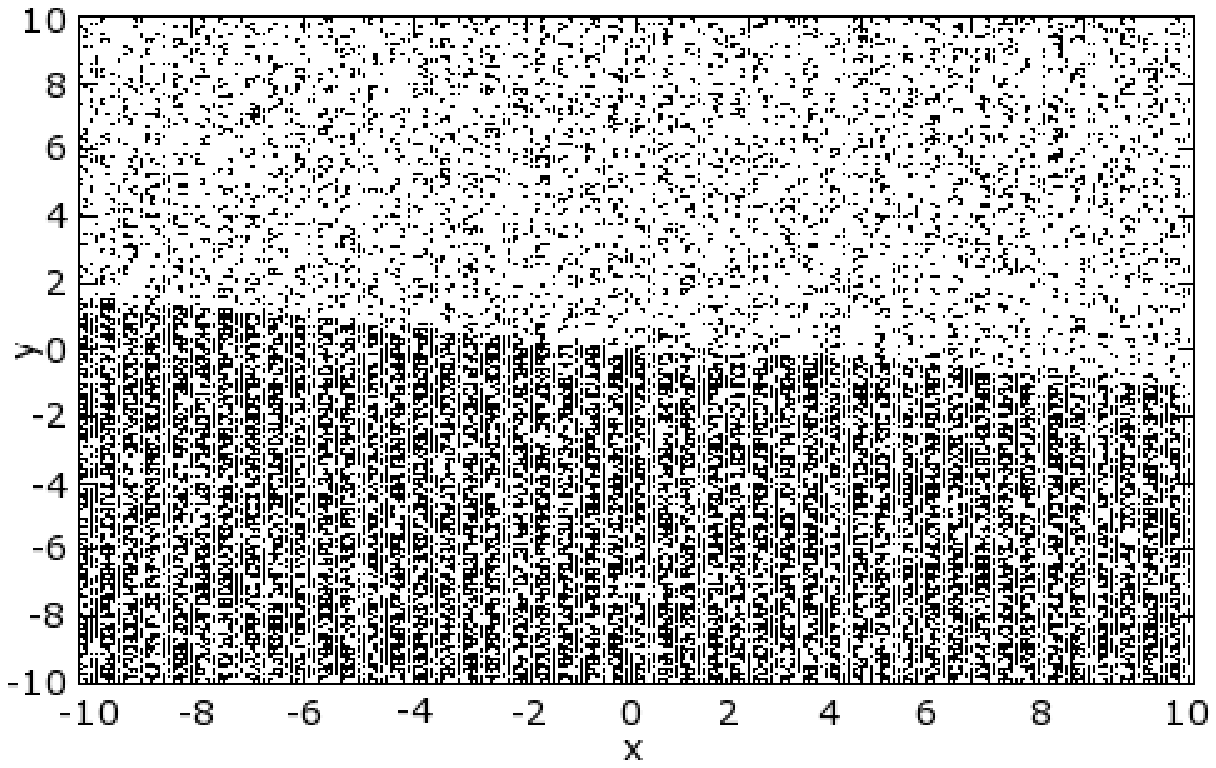,width=.5\columnwidth}}}
\caption{The basin of attraction for the different dynamical states inside the
4 cycle window. Here black region corresponds to the basins of the stable 4 
cycle as the $\mu$ value increases from intermittency towards periodic cycle.
a) $\mu=-0.0815$
b) $\mu=-0.081$
c) $\mu=-0.08$. 
In (d), the black region corresponds to bounded chaos while white region that 
of escape for $\mu=1.01$.}\label{Gu-fig6}
\end{figure}

The evolution of the basin structure in the phase space during these 
transitions is also studied in detail and shown in Fig~\ref{Gu-fig6}a--c.

The last figure~\ref{Gu-fig6}d gives the basin boundary between bounded and 
escape regions, where escape takes place beyond $\mu=+1$

\section{Synchronisation and Control of Chaos.}\label{Gu-sec5}
We try a linear mutual coupling of two GM maps as a control mechanism, by which
chaotic and  intermittency regions can be targeted to stable low periodic 
regions.

The corresponding equations are
\begin{align}\label{Gu-eq5.1}
x1_{n+1} &= y1_n+a(1-by1^2_n)y1_n+f(x1_n)+\varepsilon x2_n\notag\\[5pt]
y1_{n+1} &=-x1_n+f(x1_{n+1})\notag\\[5pt]
x2_{n+1} &= y2_n+a(1-by2^2_n)y2_n+f(x2_n)+\varepsilon x1_n\notag\\[5pt]
y2_{n+1} &=-x2_n+f(x2_{n+1})
\end{align}
where $\varepsilon$ is the coupling parameter. We analyse this system 
numerically for $\mu=-0.31$ and $a=0.008$,
$b=0.05$, both the maps individually exhibit chaotic behaviour. After coupling
with $\varepsilon=1$ and initial values $x1_0=0.1$, $y1_0=0$ for the first map 
and $x2_0=0.2$ and $y2_0=0.1$ for the second map, both the maps are found to 
settle to 5 cycles. Here control of chaos and periodicity are achieved
eventhough synchronisation is absent. 

\begin{figure}[b]
\centering
\mbox{
\subfigure[The synchronised 4 cycle of two coupled GM maps with linear and
mutual coupling for coupling parameter $\varepsilon=0.7$, 
in eqn.~\eqref{Gu-eq5.1}.]{\epsfig{file=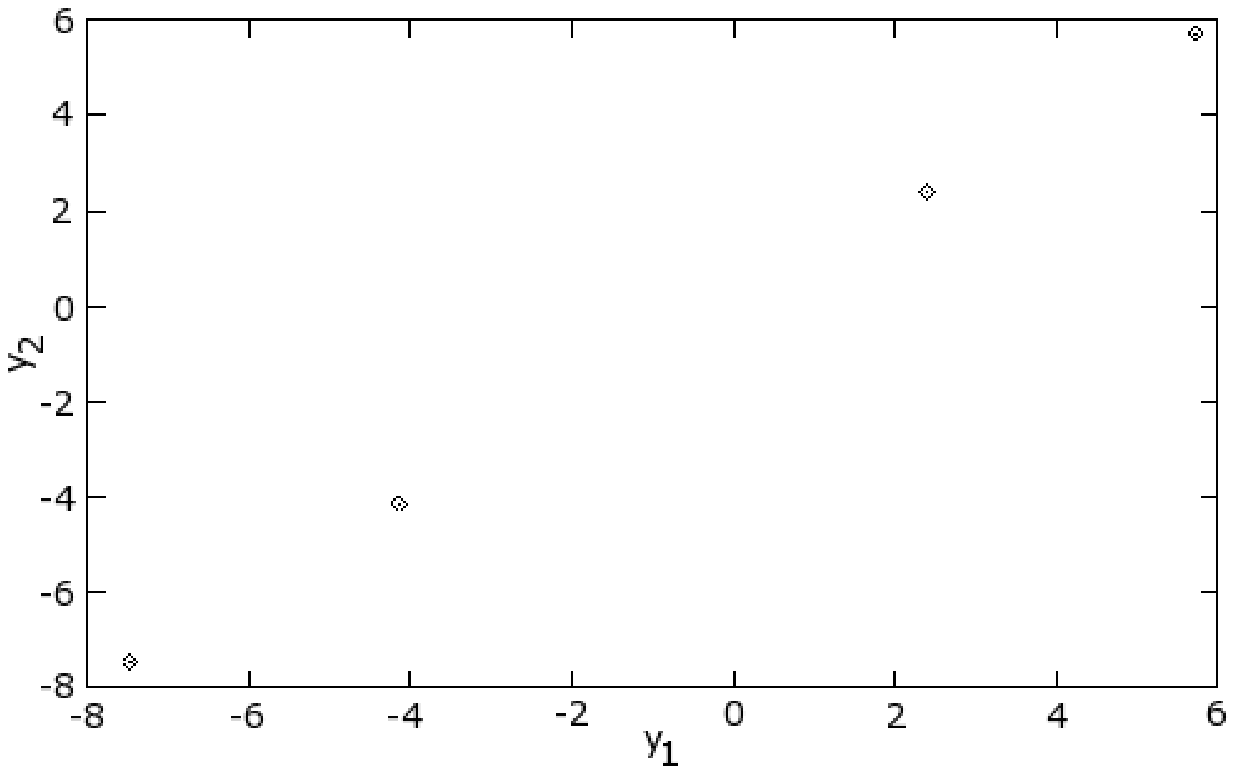,width=.5\columnwidth}}
\subfigure[The synchronisation time $\tau_1$ and the stabilisation time after
perturbation $\tau_2$ are shown in the $e^x-n$ plot of the coupled 
system]{\epsfig{file=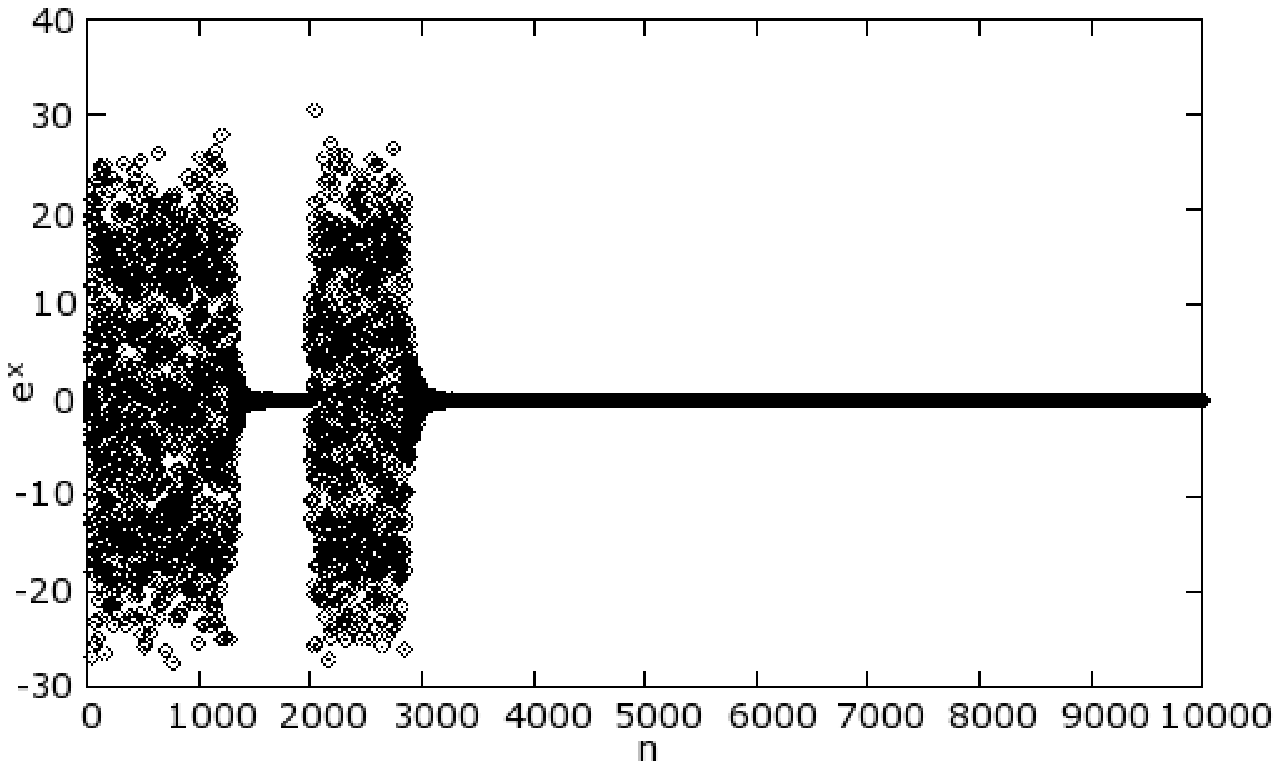,width=.52\columnwidth}}}
\caption{}\label{Gu-fig7}
\end{figure}

We are also able to synchronise two chaotic maps to  lower periodicities
using this coupling scheme. Eventhough much work is reported in the area of
synchronisation~\cite{Gumo8}, majority of these works are in continuous 
systems. However there are a few specific cases reported in the context of 
two dimensional map~\cite{Gumo9,Gumo10}.
Here we achieve synchronisation in two coupled discrete maps of the GM
type for $\mu=-0.39$ with $\epsilon=0.7$. Total synchronisation is seen in both
$x$ and $y$ as the coupled system settles to identical 4 cycles. This is 
shown in Fig~\ref{Gu-fig7}a.

The analysis is continued by developing  the error function dynamics by 
defining the error in the $x$ and $y$ values of the  two synchronisating 
systems  as $e^x=(x1-x2)$ and $e^y=(y1-y2)$. Then their  dynamics develops
through the following set of maps.
\begin{align}\label{Gu-eq5.2}
e^x_{n+1} = &\left[\mu+\frac{2(1-\mu)(x1_n+x2_n)}{(1+x1^2_n)(1+x2^2_n)}
             -\varepsilon\right]e^x_n\notag\\
            &+\left[1+a-ab(y1^2_n+y1_ny2_n+y2^2_n)\right]e^y_n\notag\\
e^y_{n+1} = &-e^x_n+\left[\mu+\frac{2(1-\mu)(x1_{n+1}+x2_{n+1})}{(1
+x1^2_{n+1})(1+x2^2_{n+1})}\right]e^x_{n+1}
\end{align}

The set of equations~\eqref{Gu-eq5.1} and~\eqref{Gu-eq5.2} are evolved together
until $e^x\rightarrow 0$ and $e^y\rightarrow 0$~\cite{Gumo11}, when total
synchronisation is said to be achieved.

From this the time for synchronisation is computed for 10 sets of initial 
values and the average is found to be $\langle\tau_1\rangle=502$ for the 
typical values mentioned above.

The robustness of the mechanism is tested by applying a perturbation and the 
stabilisation time is calculated as the time to reach synchronisation again
after it is disturbed. This is also repeated for 10 initial values
and the average time $\langle\tau_2\rangle$ is found to be 473.

Fig~\ref{Gu-fig7}b is the $e^x-n$ plot showing $\tau_1$ and $\tau_2$ values 
for a perturbation of $x_n=x_n+10$ given after 2000 iterations. A similar plot 
is obtained for the error function in $y$ also.

\section{Conclusion}\label{Gu-sec6}
In this work we analyse the dynamics of the large variety of interesting and 
lively patterns
exhibited by the GM map and find that they are attractors of the system in the 
phase plane in the neighbourhood of higher order cycles. The temporal
behaviour then corresponds to intermittency of Type I with an exponent
$\sim -0.5$.

The dependence of the phase space structure of the patterns on minute changes 
in the 
parameter makes it a useful tool in decision making algorithms. This sensitive
dependence is due to the recurring periodic and self similar substructures in
the bifurcation scenario, each with its own intermittency, periodicity,
quasi periodic band and merging of bands leading to chaos. The nature and 
characterisation of these patterns and occurrence of crises related phenomena
near them are being studied and will be reported else where.

The coupled states of these maps are either synchronised or they stabilise to
attractors of the same periodicity, without amplitude synchronisation. 
A detailed analysis of the error function dynamics is carried out to estimate 
average time for synchronisation and stabilisation time after applying a
perturbation.

The relative efficiency of different schemes of coupling in synchronising 2
such maps to chosen dynamical states is intended as further study.

\subsection*{Acknowledgement}
We thank Prof. Y. Tanaka, College of Education, Ibaraki University, Mito, Japan
for useful discussions through a private communication. One of us
(G. Ambika) thank IUCAA, Pune for hospitality and computer facility.

\clearpage

\end{document}